\documentclass[preprint,12pt]{elsarticle}

\usepackage[super]{cite}
\usepackage{xcolor}
\usepackage{graphicx}
\usepackage{amsmath,amssymb}
\usepackage[verbose,hypertexnames=false]{hyperref}

\newcommand{\Tc}{T_{\mathrm{c}}}
\newcommand{\mBP}{m^{\!\star}_{\mathrm{BP}}}

\newcommand{\nBP}{n_{\mathrm{BP}}}
\newcommand{\Om}{\Omega}

\begin{document}

\begin{frontmatter}



\title{Bipolaronic High-Temperature Superconductivity\\
from Phonon-Modulated Hopping: A Perspective}

\author{John Sous}

\address{Department of Applied Physics, Yale University, New Haven, Connecticut 06511, USA\\
Energy Sciences Institute, Yale University, West Haven, Connecticut 06516, USA\\
john.sous@yale.edu}



\begin{abstract}
Phonon-mediated superconductivity is conventionally thought to be capped at a transition temperature $\Tc$ no larger than roughly one-tenth of the phonon frequency $\Om$, a bound rooted in the breakdown of Migdal--Eliashberg theory at intermediate coupling and in the heaviness of bipolarons formed in standard models with phonons that couple to the electron density. In this review I describe a route to phonon-mediated high-$\Tc$ superconductivity that bypasses this bound. The key ingredient is a class of electron--phonon couplings in which lattice distortions modulate the electron  \emph{hopping} and therefore its kinetic energy rather than its potential energy, known as the Peierls model (also known as Su--Schrieffer--Heeger model). In these models phonon exchange generates an interaction that binds two electrons into a small but unusually \emph{light} bipolaron. Using sign-problem-free quantum Monte Carlo simulations of a bond-Peierls model on the square and cubic lattices, my collaborators and I have shown that a dilute liquid of such bipolarons forms an $s$-wave superconductor with a $\Tc/\Om$ that significantly exceeds the conventional bound, that this conclusion is robust against screened Coulomb repulsion, and that $\Tc/\Om$ --- despite being reduced --- remains above bound in presence of strong long-range Coulomb repulsion.  A semi-classical instanton analysis explains why, at strong coupling, bipolarons in models with phonon-modulated hopping are lighter than their density-coupled (Holstein) counterparts. I close with a discussion of materials in which this physics may be operative, in particular the iron-based pnictide superconductors, and of design principles that follow from it. This is a proceedings contribution to the Athens Workshop in Theoretical Physics: 10th Anniversary, held at the National and Kapodistrian University of Athens on December 17--19, 2025.
\end{abstract}


\end{frontmatter}

\tableofcontents

\section{Introduction}

The problem of how high a transition temperature $\Tc$ can be achieved by phonon-mediated superconductivity is more than half a century old, and it is still open. In this review I want to describe a particular line of attack on this question that my collaborators and I have pursued over the past several years. The starting point is to take the strong-coupling, low-density regime seriously and to ask: what kinds of electron--phonon couplings actually allow it to support a high-$\Tc$ superconductor, and which kinds do not? The answer turns out to depend, in a strikingly sharp way, on whether the phonons modulate the electron \emph{density} or the electron \emph{hopping}.

The conventional picture is by now standard textbook material. In the weak-coupling, adiabatic regime, the Bardeen--Cooper--Schrieffer (BCS) theory of phonon-induced pairing\cite{BCS} and its strong-coupling extension by Migdal and Eliashberg\cite{Migdal,Eliashberg,Bergmann73} describe how a Fermi liquid of electrons becomes unstable to a Cooper-paired condensate at a $\Tc$ that is exponentially small in the inverse dimensionless electron--phonon coupling $\lambda$, and that grows as $\lambda$ is increased. As $\lambda$ approaches values of order unity, however, the Migdal--Eliashberg expansion breaks down: bipolarons and lattice instabilities preempt the Fermi-liquid description.\cite{Alexandrov2001,WernerMillis,BauerHanGunnarsson,EsterlisPRB2018,EsterlisNPJ2018,EsterlisTcBound} The empirical maximum of $\Tc$ as a function of $\lambda$ is reached roughly where this breakdown sets in, and one is led to a rough but widely accepted bound, $\Tc \lesssim \Om/10$, on the transition temperature attainable from phonon-mediated mechanisms at fixed phonon frequency. For typical phonon energies $\Om \sim 300~\mathrm{K}$, this gives $\Tc$ of order $30~\mathrm{K}$, the well-known ``conventional ceiling'' that the high-pressure hydrides circumvent only by raising $\Om$ itself.\cite{H3S1,H3S2}

The complementary, strong-coupling perspective takes the dilute limit of preformed pairs as its starting point. Here electrons are imagined to be tightly bound by phonon exchange into local singlet bipolarons; the superconductor is then a Bose condensate (or, in two dimensions, a Berezinskii--Kosterlitz--Thouless superfluid) of these objects, with $\Tc$ controlled by the bipolaron density and the inverse of their effective mass. Within the standard model of strong electron--phonon coupling --- the Holstein model, in which phonons couple to the local electron density --- this route was concluded already in the 1990s to be a dead end: bipolarons in the Holstein model are exponentially heavy at strong coupling, the Bose condensation temperature is therefore vanishingly small, and high-$\Tc$ bipolaronic superconductivity was deemed impossible.\cite{ChakravertyBipolaron}

The point I want to make in this review is that this last conclusion, while correct for density-coupled (Holstein) models, is \emph{not} a property of bipolarons in general. In the family of phonon-modulated-hopping models that includes the Su--Schrieffer--Heeger (SSH) and Peierls models, bipolarons are bound in a qualitatively different way --- through a phonon-mediated, kinetic-energy-enhancing pair-hopping interaction --- and they remain comparatively \emph{light} even at strong coupling. As a result, these models support an $s$-wave bipolaronic superconductor whose $\Tc$, when computed numerically exactly, can be substantially larger than the rough bound $\Om/10$ over a wide swath of parameter space, can survive both a large local Hubbard $U$ and an unscreened long-range Coulomb tail, and can exceed in the relevant regime any value attainable from Migdal--Eliashberg theory of strong-coupling superconductivity in the same model. The mechanism is, to my knowledge, the only one currently known in which a physically reasonable form of phonon-mediated coupling generates a bipolaronic high-$\Tc$ superconductor without invoking large phonon frequencies.

This paper is adapted from talks delivered at seminars and workshops. I begin in Sec.~\ref{sec:weak} with a brief recap of weak-coupling phonon-mediated superconductivity, the Migdal--Eliashberg framework, and the bound $\Tc \lesssim \Om/10$ that follows from its analysis. Section~\ref{sec:bipolarons} reviews why bipolarons are heavy in density-coupled models and why this was historically taken to rule out bipolaronic high-$\Tc$ superconductivity. Section~\ref{sec:peierls} introduces phonon-modulated-hopping (SSH/Peierls) models and presents the original observation\cite{SousPRL2018} that, in the antiadiabatic limit, they support light bipolarons through phonon-mediated electron-pair hopping. Section~\ref{sec:bond2D} turns to the bond-Peierls variant of the model, which is amenable to sign-problem-free quantum Monte Carlo, and reviews the resulting phase diagram of high-$\Tc$ bipolaronic superconductivity in two dimensions in the presence of a screened Hubbard repulsion.\cite{ZhangSousPRX2023} Section~\ref{sec:bond3D} extends the analysis to three dimensions and to unscreened long-range Coulomb repulsion.\cite{SousZhangPRB2023} Section~\ref{sec:semiclassical} describes a recent semi-classical instanton-based theory\cite{KimHanSousPRB2024} that explains, asymptotically exactly, \emph{why} bipolarons in phonon-modulated-hopping models are lighter than their density-coupled counterparts. I conclude in Sec.~\ref{sec:outlook} with a discussion of materials, design principles, and open directions.

I have tried to write this review for a general physics audience: condensed-matter readers should not find anything missing, but I hope that high-energy and AMO readers will be able to follow without difficulty as well. Throughout I set $\hbar = k_{\mathrm{B}} = 1$.

\section{Weak-coupling superconductivity and the Migdal--Eliashberg ceiling}\label{sec:weak}

I will begin with a few words on the conventional picture, more to fix language than to review what is in the textbooks. In an ordinary metal, electrons repel one another, but a weak coupling to lattice vibrations dresses each electron in a slowly responding cloud of ionic distortions and induces, on long time scales, an effective retarded attraction between electrons mediated by the exchange of phonons. The Fermi liquid is unstable to this attraction: a Cooper instability sets in, the ground state acquires a macroscopic occupation of the paired channel with a pairing gap $\Delta$, and the system becomes a superconductor --- a dissipationless conductor at sufficiently low temperatures.\cite{BCS} The BCS transition temperature is exponentially small in $1/\lambda$, where $\lambda$ is the dimensionless electron--phonon coupling, and grows monotonically as $\lambda$ is increased.

A natural question is what happens when $\lambda$ is no longer small. The standard answer is provided by Migdal--Eliashberg (ME) theory,\cite{Migdal,Eliashberg,Bergmann73} which is most cleanly motivated as an expansion in the small parameter $\Om/E_{\mathrm{F}}$, where $E_{\mathrm{F}}$ is the Fermi energy. Within this expansion vertex corrections are formally suppressed by powers of $\lambda(\Om/E_{\mathrm{F}})$, and one obtains a tractable set of self-consistent equations for the electron self-energy and the anomalous (gap) function. Solving these equations,\cite{Bergmann73} one finds that $\Tc$, expressed in units of the renormalized phonon frequency, evolves smoothly from the BCS form $e^{-1/\lambda}$ at small $\lambda$ to the Allen--Dynes asymptote $\sqrt{\lambda}$ at large $\lambda$, and McMillan's phenomenological fit to experimental data on conventional superconductors recovers this behavior with empirical Coulomb-pseudopotential corrections.\cite{McMillanFormula}

\begin{figure}[t]
\centerline{\includegraphics[width=0.8\columnwidth]{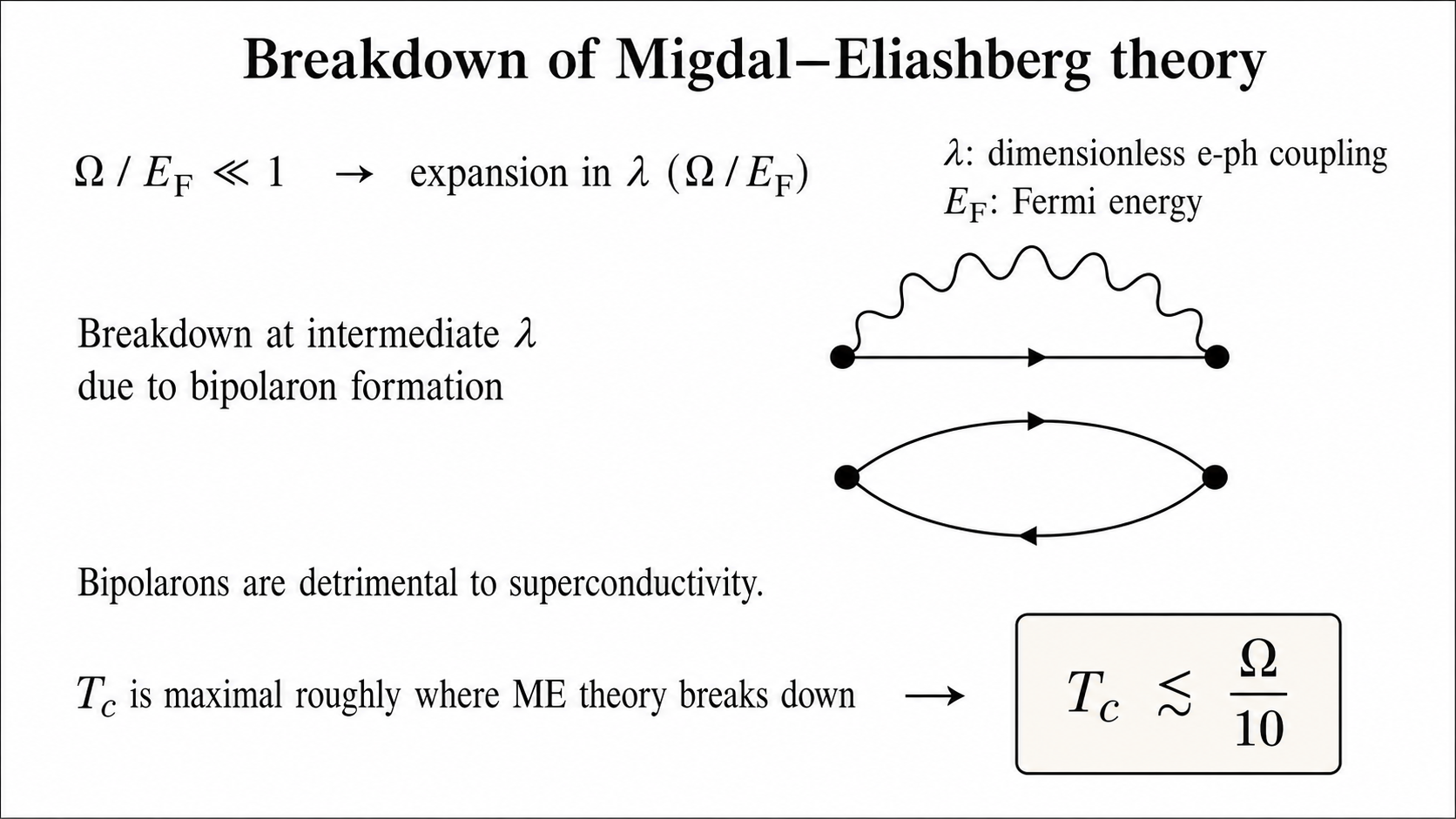}}
\caption{Schematic illustration of the breakdown of Migdal--Eliashberg theory at intermediate $\lambda$ and the resulting bound $\Tc \lesssim \Om/10$ on phonon-mediated superconductivity out of a Fermi liquid.}
\label{fig:ME}
\end{figure}

The crucial subtlety is that the ME framework does not, by itself, know about polarons or bipolarons. As the bare coupling is increased, two things happen that are invisible to ME theory but very visible to a more honest treatment of the problem. First, the Fermi-liquid state becomes a metastable state, higher in energy than a state of bipolarons, beyond a critical coupling of order unity.\cite{Alexandrov2001,WernerMillis,BauerHanGunnarsson,EsterlisPRB2018,EsterlisNPJ2018,EsterlisTcBound} Second, the lattice itself can become unstable, either to a structural reconstruction or to a charge-density-wave order whose onset reduces the effective coupling.\cite{CohenBounds,EsterlisTcBound} As a result, increasing $\lambda$ beyond $\lambda \sim 1$ does not buy any further increase in $\Tc$; the maximum $\Tc/\Om$ accessible to phonon-mediated superconductivity out of a Fermi liquid is reached very near where ME theory breaks down. Various careful analyses of this point yield, as a rough but robust statement,
\begin{equation}
\Tc \;\lesssim\; \frac{\Om}{10}.
\label{eq:McMillanBound}
\end{equation}
For a typical phonon energy $\Om \sim 300~\mathrm{K} \approx 0.03~\mathrm{eV}$, Eq.~\eqref{eq:McMillanBound} translates into $\Tc \lesssim 30~\mathrm{K}$. This is the well-known ``conventional ceiling.'' The high-pressure hydrides circumvent it not by violating the bound, but by raising $\Om$.\cite{H3S1,H3S2} This weak-coupling logic and its breakdown are summarized schematically in Fig.~\ref{fig:ME}.

I want to underline two implicit assumptions hidden in the chain of reasoning that leads to Eq.~\eqref{eq:McMillanBound}. The first is that the Fermi-liquid/ME description is the correct starting point at intermediate to strong coupling. The second is that, when bipolarons do form, they must be heavy and detrimental to superconductivity.
Both assumptions are defensible in parts of the conventional density-coupled literature, but for different reasons and in different regimes. The Fermi-liquid/ME starting point is appropriate at sufficiently high density or weak to moderate coupling, but its limitations at low density and strong coupling are already well known in the Holstein problem~\cite{Alexandrov2001, WernerMillis, EsterlisPRB2018}. The expectation of heavy bipolarons, by contrast, is characteristic of the local Holstein limit. In phonon-modulated-hopping models these two pieces of intuition should therefore be disentangled: the low-density problem is not naturally organized around a Fermi-liquid expansion, as is already familiar from density-coupled models, but the bipolarons that form need not be heavy.
\section{Bipolarons and why they are heavy in density-coupled models}\label{sec:bipolarons}

\begin{figure}[h]
\centerline{\includegraphics[width=0.8\columnwidth]{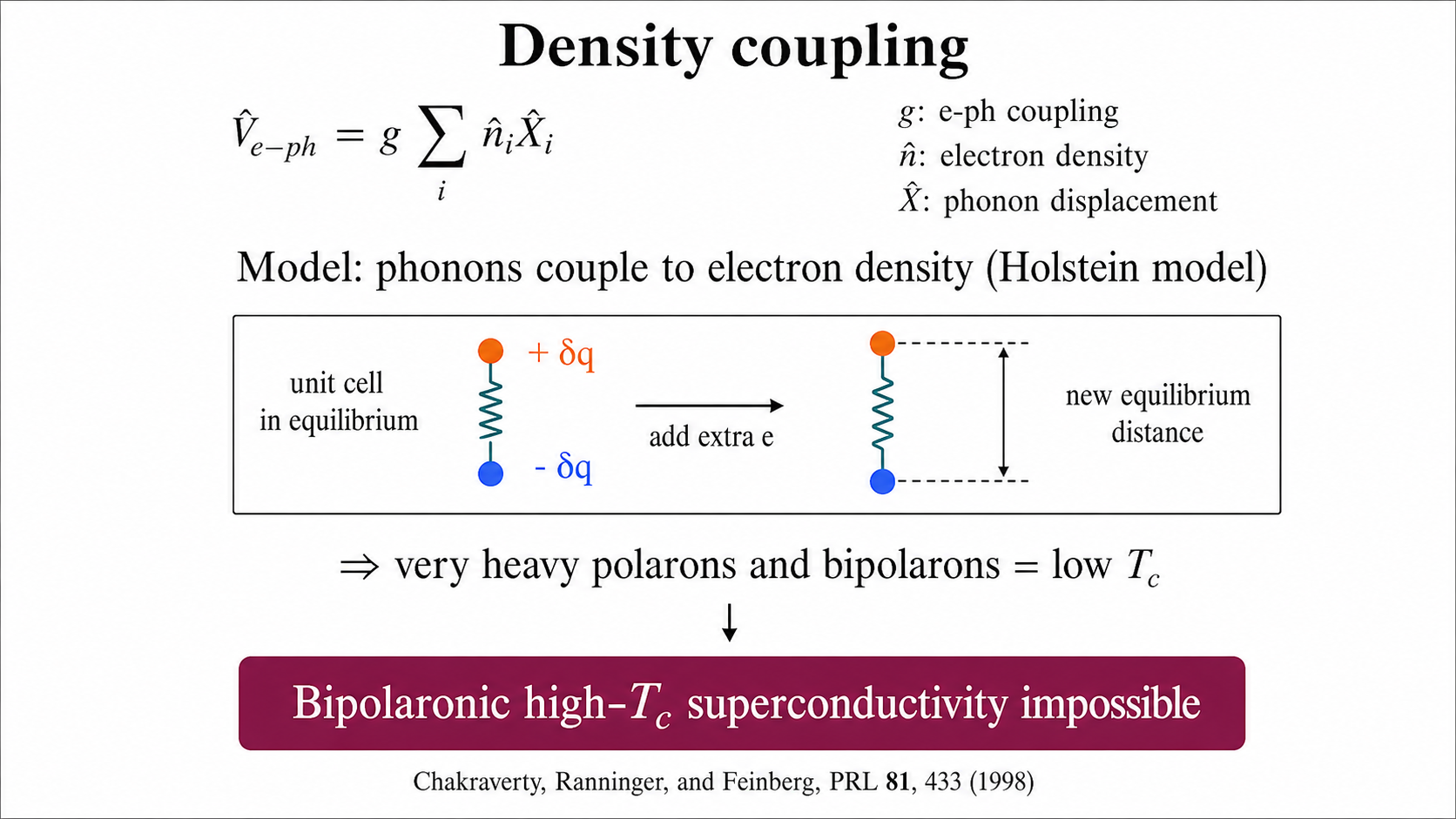}}
\caption{Density coupling: a phonon $\hat{X}_i$ attached to each lattice site couples to the local electron density $\hat{n}_i$. Adding an electron displaces the oscillator and digs a deep self-trapping well; the resulting polaron, and the bipolaron formed from two of them, is heavy.}
\label{fig:Holstein}
\end{figure}

Let me now switch to the strong-coupling, low-density perspective. The basic objects are polarons --- electrons dressed by clouds of phonons --- and bipolarons --- bound singlet pairs of polarons held together by phonon exchange. In the dilute limit the bipolarons are well-defined particles, and their collective state is a Bose superfluid (in two dimensions a Berezinskii--Kosterlitz--Thouless superfluid; in three dimensions a Bose--Einstein condensate). The transition temperature scales as the inverse of the bipolaron mass, so the entire question of whether bipolaronic superconductivity is competitive reduces, in the dilute limit, to: how heavy is a bipolaron?

The conventional answer was articulated cleanly by Chakraverty, Ranninger, and Feinberg\cite{ChakravertyBipolaron}. The intuition is straightforward, and it is correct \emph{within the model in which it is given.} The model in question is the Holstein model, in which a local oscillator $\hat{X}_i$ on each lattice site couples to the local electron density $\hat{n}_i$:
\begin{equation}
\hat{V}_{\mathrm{e\text{-}ph}}^{\mathrm{H}} \;=\; g \sum_i \hat{n}_i \hat{X}_i.
\label{eq:Holstein}
\end{equation}
Adding an electron to a site shifts the equilibrium distortion of the local oscillator; the electron now sits in a deeper potential well dug by its own coupling, and to move to a neighboring site it must drag this distortion with it. In the strong-coupling limit the polaron's effective mass grows exponentially in $\lambda$, and the bipolaron, which is a bound pair of two such heavy objects, is even heavier --- its mass enhancement is roughly the square of the polaron's. The corresponding Bose superfluid temperature is then exponentially suppressed, and one is forced to the conclusion of Ref.~\cite{ChakravertyBipolaron}: bipolaronic high-$\Tc$ superconductivity is impossible. 
The conventional density-coupled picture reviewed is sketched in Fig.~\ref{fig:Holstein}.

I want to emphasize that this argument rests entirely on a particular form of the electron--phonon coupling: the phonon talks to the local density, and a self-trapping potential well is the result. Under this assumption the conclusion is unimpeachable. The question I want to ask, and the question that organizes the rest of this review, is whether the same conclusion follows when the phonon couples not to the density but to the hopping.

\section{Phonon-modulated hopping: light bipolarons in the intermediate antiadiabatic limit}\label{sec:peierls}

The model that motivates this question is the Su--Schrieffer--Heeger (SSH) model, originally introduced to describe conducting polymers,\cite{SSH} or, in the closely related lattice form, the Peierls model.\cite{BarisicPeierls1} In its site-Peierls version one places an oscillator $\hat{X}_i$ on each lattice site, but couples it to the electron \emph{hopping} amplitude rather than to the local density:
\begin{equation}
\hat{V}_{\mathrm{e\text{-}ph}}^{\mathrm{sP}} \;=\; g \sum_{\langle i,j\rangle,\sigma} \big(\hat{c}^\dagger_{i,\sigma}\hat{c}_{j,\sigma} + \mathrm{h.c.}\big)\big(\hat{X}_i - \hat{X}_j\big).
\label{eq:sitePeierls}
\end{equation}
Physically, this is what one obtains when the relative displacement of two ions modulates the barrier (and therefore the matrix element) for an electron to hop between sites localized on those ions. It is the natural form of electron--phonon coupling in any system with more than one atom per unit cell, and it has long been argued as possibly appropriate for the description of buckled bonds as in the cuprates and of analogous geometries in many transition-metal oxides.\cite{ZXNagaosaPeierlsCuprates,90bond}

Single polarons in the model of Eq.~\eqref{eq:sitePeierls} were already known to be qualitatively different from their Holstein counterparts.\cite{Marchand2010,CarboneMillisReichmanSousBondPolaron,ZhangProkofevSvistunovBondPolaron} In particular, a sharp polaronic transition appears at a critical coupling $\lambda_{\mathrm{c}}$ above which the polaron \emph{lightens} as the coupling is increased --- the opposite of the Holstein behavior --- because the dressing converts the bare nearest-neighbor hopping into a longer-ranged, sign-alternating hopping.\cite{Marchand2010} The natural follow-up question --- what does this mean for two electrons? --- was the subject of Ref.~\cite{SousPRL2018}, on which I will focus in this section.

\begin{figure}[t]
\centerline{\includegraphics[width=0.525\columnwidth]{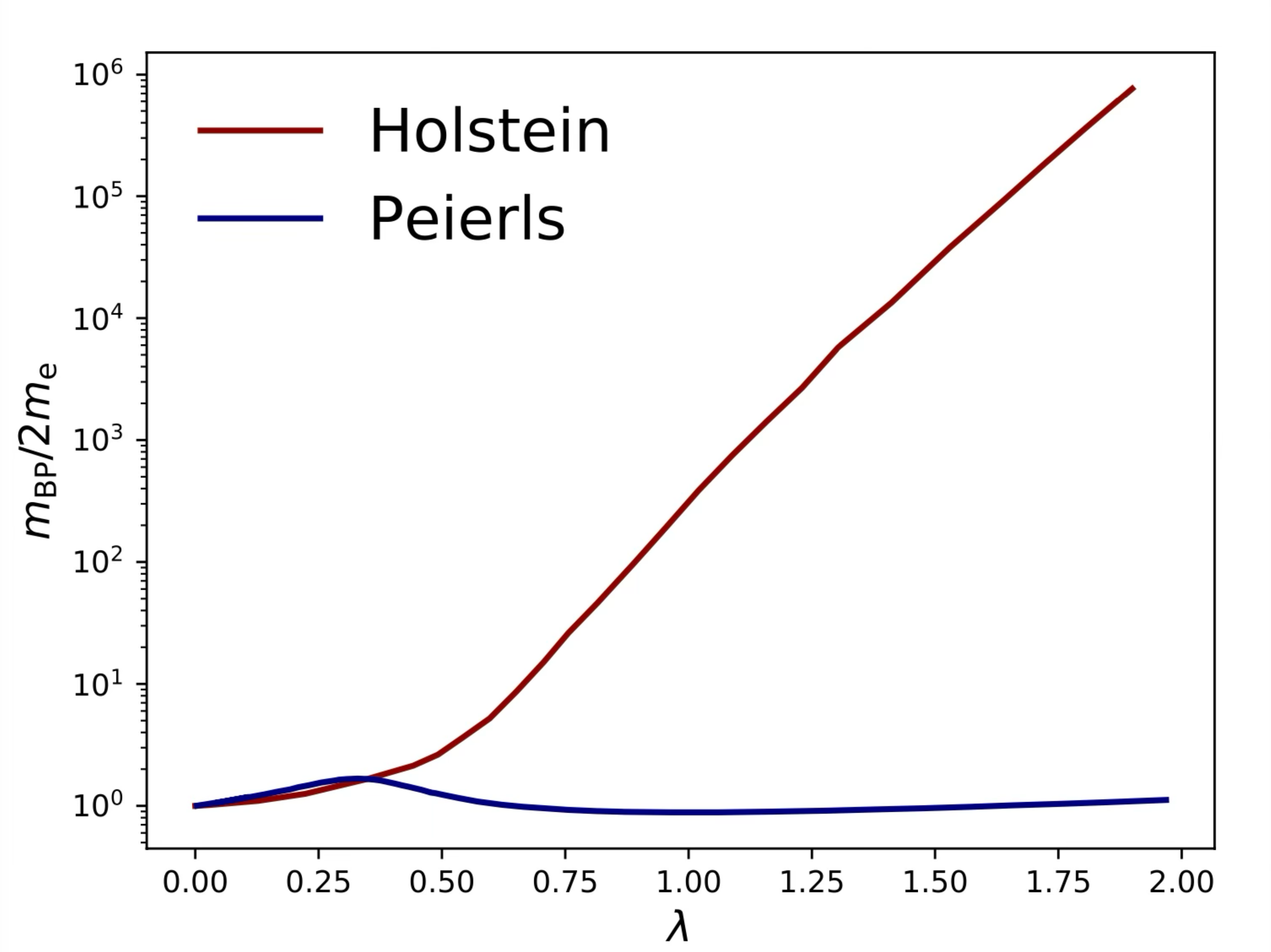}}
\vspace{-2mm}
\caption{Bipolaron effective mass $\mBP/(2m_{\mathrm{e}})$ in the one-dimensional site-Peierls model (blue) compared with the Holstein model (red), as a function of the dimensionless electron--phonon coupling $\lambda$ at $\Om \sim W$. The Peierls bipolaron remains light up to and beyond $\lambda = 2$, while the Holstein bipolaron is exponentially heavy. Adapted from Ref.~\protect\cite{SousPRL2018}.}
\label{fig:Peierls1D}
\end{figure}

In Ref.~\cite{SousPRL2018}, my collaborators Chakraborty, Krems, Berciu and I solved the two-electron singlet sector of Eq.~\eqref{eq:sitePeierls} on a one-dimensional chain in the regime $\Om \sim W$ (with $W$ the bare bandwidth) using a numerically exact\cite{ CarboneReichmanSousPRB2021} Green-function approach.\cite{BerciuPRL2006} The result is shown schematically in the inset of Fig.~\ref{fig:Peierls1D}. The bipolaron mass $\mBP$, expressed in units of twice the bare electron mass $m_{\mathrm{e}} = 1/(2t)$, grows by less than an order of magnitude over the entire range of couplings up to $\lambda = 2$, in stark contrast to the Holstein case, where it grows by several orders of magnitude over the same range. The bipolaron is, by any reasonable standard, light.

\begin{figure}[h]
\centerline{\includegraphics[width=0.85\columnwidth]{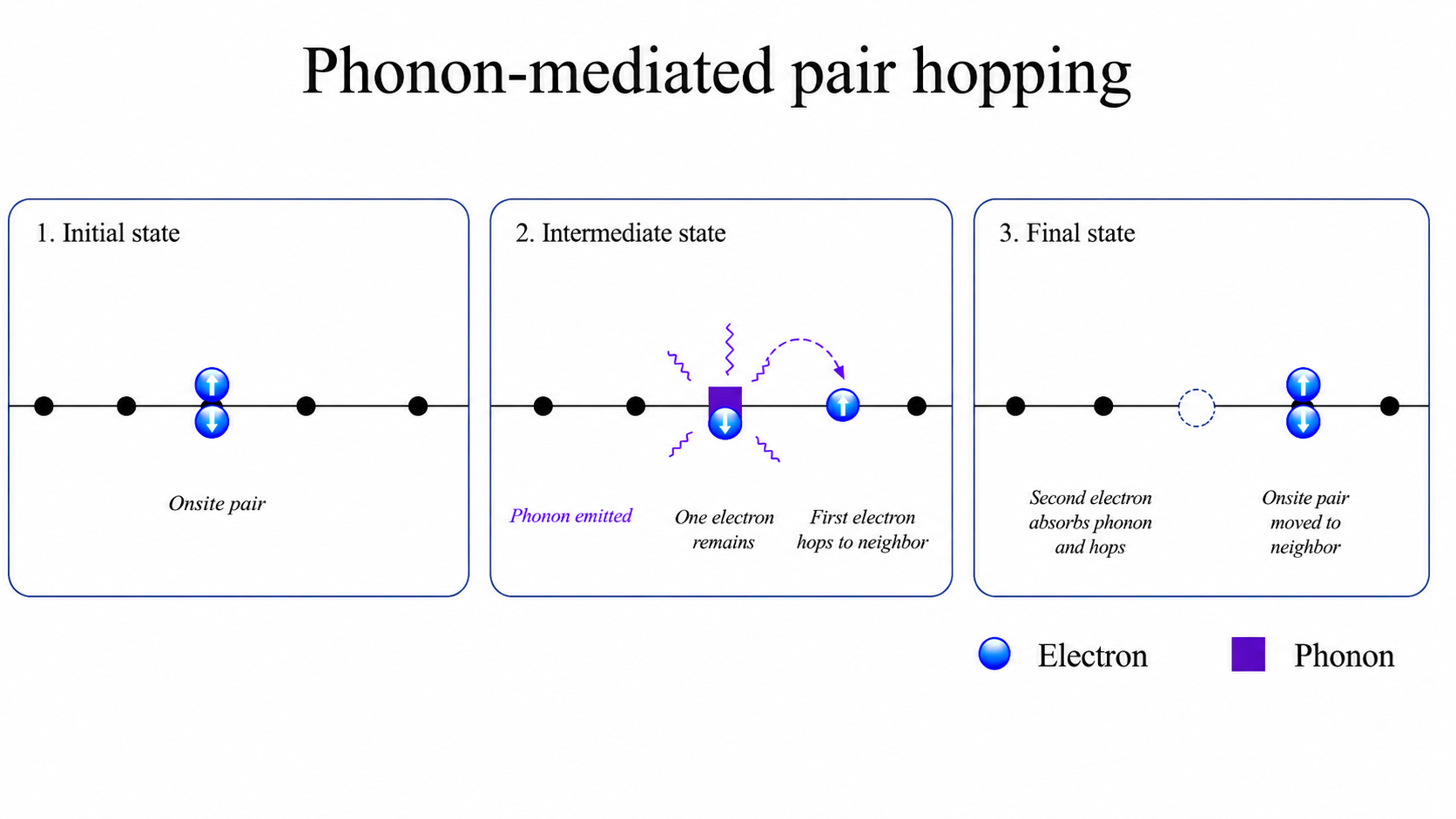}}
\vspace{-7mm}
\caption{Cartoon of the phonon-mediated electron-pair-hopping process generated by the site-Peierls coupling of Eq.~\eqref{eq:sitePeierls} in the antiadiabatic limit $\Om/t\gg 1$. A virtual phonon emitted by one electron during a hop is absorbed by a neighboring electron during a second hop, with the net effect of moving a singlet pair as a unit.}
\label{fig:pairhop}
\end{figure}

The mechanism behind this result is clean and worth describing in some detail, because it is the same mechanism that is at work in most of that follows. In the antiadiabatic limit $\Om/t \gg 1$, second-order perturbation theory in the electron--phonon coupling generates an effective electron--electron interaction by integrating out a single virtual phonon. For the site-Peierls coupling of Eq.~\eqref{eq:sitePeierls}, the leading process is depicted in Fig.~\ref{fig:pairhop}: an electron on site $j$ creates a virtual phonon while hopping to site $i$, and a second electron with an opposite spin on site $j$ then absorbs the same phonon while hopping to its partner's site, site $i$. The net result is that a pair of electrons hops, as a unit, to a neighboring site. The induced interaction is, in the simplest version, a pair-hopping term of the form
\begin{equation}
\hat{H}_{\mathrm{pair}}^{\mathrm{eff}} \;\sim\; -\frac{2g^2}{\Om - U}\sum_{\langle i,j\rangle}\big(\hat{c}^\dagger_{i\uparrow}\hat{c}^\dagger_{i\downarrow}\hat{c}_{j\downarrow}\hat{c}_{j\uparrow} + \mathrm{h.c.}\big),
\label{eq:pairhop}
\end{equation}
where $U$ is the on-site Hubbard repulsion. This is a kinetic-energy-enhancing interaction: a bound singlet pair lowers its energy by hopping coherently between neighboring bonds, and the effective dynamics it generates for the pair is what gives the bipolaron its light mass. It is the absence of any analogous process in the Holstein model --- where phonon exchange instead generates an instantaneous on-site attraction $-g^2/\Om \sum_i \hat{n}_{i\uparrow}\hat{n}_{i\downarrow}$ that has nothing to do with the kinetic energy --- that makes the bipolaron heavy in that case.

Two limitations of the analysis of Ref.~\cite{SousPRL2018} should be acknowledged at this point, since they are precisely the limitations that the work I describe in the next two sections was designed to overcome. The first is that the analysis was restricted to a single spatial dimension, and the regime $\Om \sim W$ studied there is not the regime relevant to most real materials, where one expects $\Om \ll W$. The second is more serious: the work computed bipolaron properties but did not address superconductivity directly. To say something about $\Tc$, one needs not only the mass of a single bipolaron but also a controlled estimate of the density at which a dilute liquid of bipolarons condenses, and one needs both quantities in two or three spatial dimensions, in regimes of $t/\Om$ relevant to real materials, and ideally in the presence of a realistic Coulomb repulsion. None of this was accessible to the methods of Ref.~\cite{SousPRL2018}. The next sections describe how this gap was closed.

\section{Bond-Peierls bipolaronic high-$T_{\mathrm{c}}$ superconductivity in two dimensions}\label{sec:bond2D}

\begin{figure}[t]
\centering
\begin{minipage}[c]{0.45\columnwidth}
  \centering
  \includegraphics[width=\linewidth]{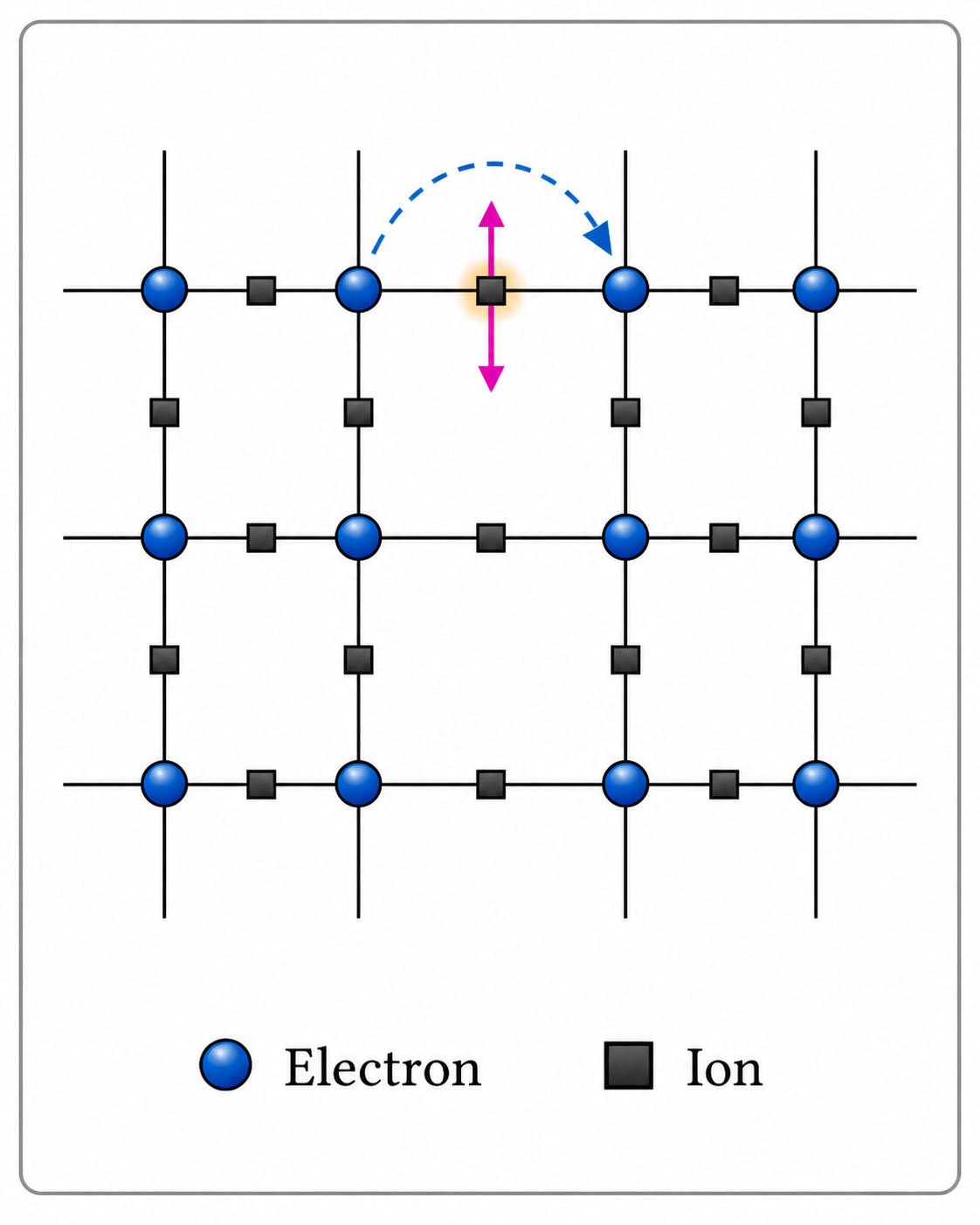}
\end{minipage}\hfill
\begin{minipage}[c]{0.45\columnwidth}
  \centering
  \includegraphics[width=\linewidth]{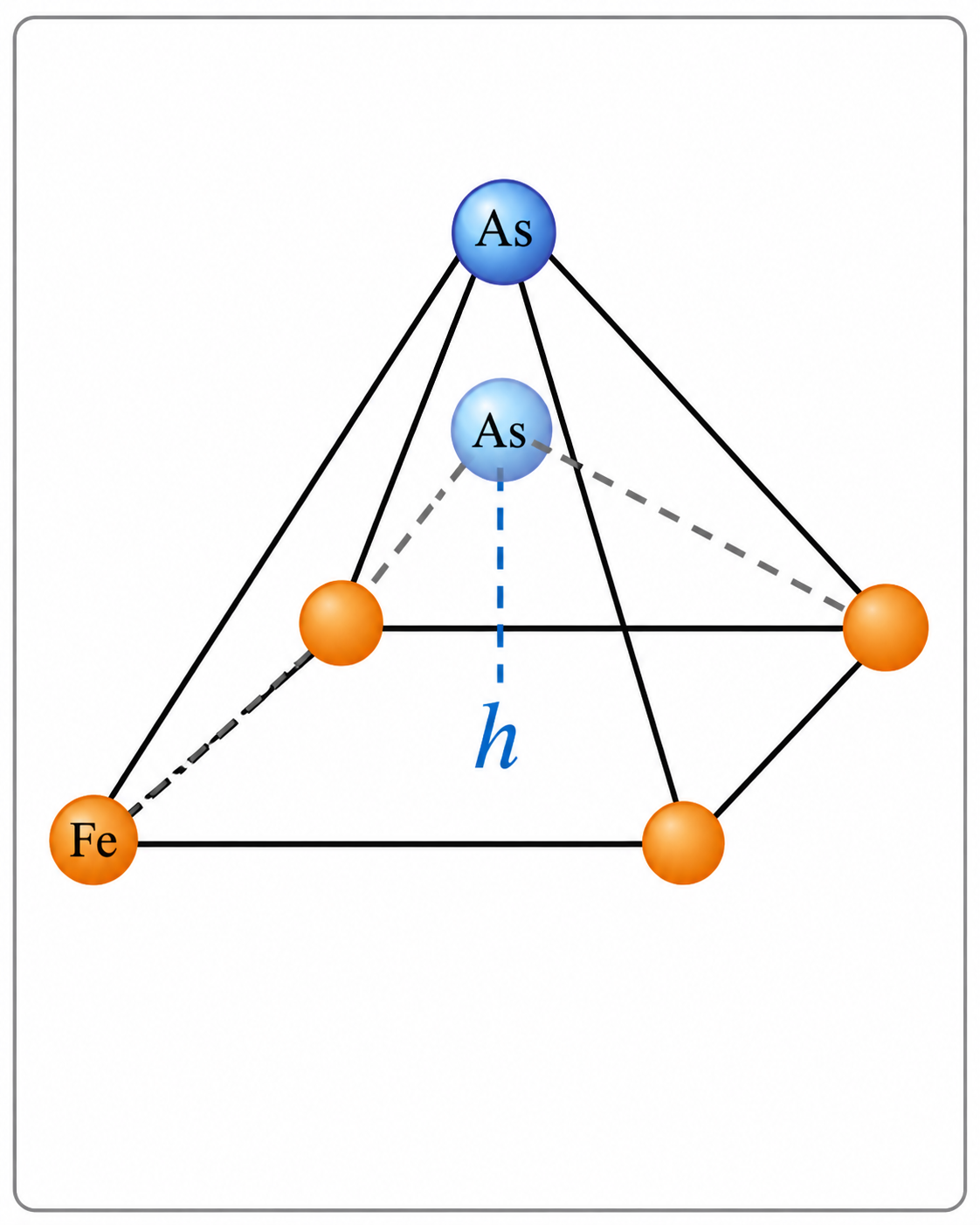}
\end{minipage}
\caption{Left: Cartoon of the bond-Peierls model: oscillators sit on the bonds of the square lattice (open ions), and their displacements $\hat{X}_{\langle i,j\rangle}$ modulate the hopping of electrons between neighboring sites (filled circles). Right: The bond-Peierls coupling in the iron pnictides. Pnictogen oscillations out of the Fe plane modulate the electron tunneling between neighboring iron atoms; destructive interference between the direct $d_{xy}$--$d_{xy}$ hop and the indirect Fe--pnictogen--Fe path produces a small net hopping with a large coupling to the pnictogen displacement. Adapted from Refs.~\protect\cite{ZhangHarriger}.}
\label{fig:bondcartoon}
\end{figure}

The model studied in this section is a close cousin of Eq.~\eqref{eq:sitePeierls} in which the oscillator that modulates the hopping between two sites lives not on either of the sites but on the \emph{bond} that connects them. Concretely, on a square lattice the electron--phonon coupling is
\begin{equation}
\hat{V}_{\mathrm{e\text{-}ph}}^{\mathrm{bP}} \;=\; \alpha\sum_{\langle i,j\rangle,\sigma}\big(\hat{c}^\dagger_{i,\sigma}\hat{c}_{j,\sigma} + \mathrm{h.c.}\big)\big(\hat{b}^\dagger_{\langle i,j\rangle} + \hat{b}_{\langle i,j\rangle}\big),
\label{eq:bondPeierls}
\end{equation}
where $\hat{b}^\dagger_{\langle i,j\rangle}$ creates a quantum of a single Einstein oscillator associated with the bond $\langle i,j\rangle$, with frequency $\Om$. The full Hamiltonian is
\begin{equation}
\hat{\mathcal{H}} \;=\; -t\sum_{\langle i,j\rangle,\sigma}\big(\hat{c}^\dagger_{i,\sigma}\hat{c}_{j,\sigma} + \mathrm{h.c.}\big) + U\sum_i \hat{n}_{i\uparrow}\hat{n}_{i\downarrow} + \Om\sum_{\langle i,j\rangle}\hat{b}^\dagger_{\langle i,j\rangle}\hat{b}_{\langle i,j\rangle} + \hat{V}_{\mathrm{e\text{-}ph}}^{\mathrm{bP}},
\label{eq:Hbond}
\end{equation}
with dimensionless coupling $\lambda \equiv \alpha^2/(2\Om t)$ and adiabaticity ratio $t/\Om$. I refer to this as the bond-Peierls model. Physically, the bond oscillator should be thought of as a ligand or apical ion sitting between the two electronic sites, whose transverse motion modulates the barrier through which the electrons tunnel; I will return to the materials motivation below. Figure~\ref{fig:bondcartoon} illustrates both the idealized bond-Peierls lattice model and the corresponding pnictide geometry.

\subsection{The bond-Peierls model and its physical motivation}

A natural setting in which the coupling of Eq.~\eqref{eq:bondPeierls} arises is one in which an out-of-plane atom sits between two in-plane sites, and its motion transverse to the bond modulates the electronic tunneling between them. This is, in particular, the situation in the iron-based pnictide superconductors,\cite{HauleNatMater,PnictogenHeight,ZXSHENFESE} where a pnictogen atom sits above or below the Fe plane and its $z$-displacement modulates two competing hopping pathways between neighboring Fe sites: a direct $d_{xy}$--$d_{xy}$ hop and a second-order superexchange-like process through the pnictogen $p_x$ orbital. The two amplitudes have opposite signs and very nearly cancel in FeSe, with the consequence that small displacements of the pnictogen produce a fractionally large modulation of the net hopping --- the physical origin of a large bond-Peierls coupling. This statement should be understood as applying to the leading linear term in an expansion of the hopping about the equilibrium geometry. In a real material the hopping is a nonlinear function of displacement and in generic situations higher-order terms become important before the linearized hopping is driven through zero. The pnictide geometry is special because the equilibrium hopping can be anomalously small as a result of cancellation between competing paths, allowing a large fractional modulation for relatively small absolute displacements.  The relevant ratio of the in-plane Fe--Fe hopping to the transverse pnictogen phonon frequency in FeSe is estimated to be $t/\Om \sim 2$--$3$,\cite{ZXSHENFESE,GerberScience} which, as I will show below, is close to the regime in which the bipolaronic mechanism is at its most effective. Variants of the same physics --- buckled-bond modulation of the Cu--O hopping --- have been discussed extensively in the cuprates by Devereaux, Shen, Zaanen, Nagaosa, Gunnarsson and others.\cite{ZXNagaosaPeierlsCuprates,90bond,corner-sharing} Pump--probe experiments on FeSe have directly measured the modulation of the iron $d$-band energies by the relevant phonon mode.\cite{GerberScience} I emphasize, however, that the point of this discussion is only to identify a physically relevant realization of a large bond-Peierls coupling. The analysis below deliberately isolates this coupling and does not include many other ingredients that are essential in realistic descriptions of the pnictides or cuprates, including multiorbital structure, magnetic correlations, longer-range interactions, and material-specific band details. Thus the results should not be read as a claim that the mechanism studied here \emph{must} be operative in these materials, but rather as evidence that variants of it are natural candidates for further investigation. In interpreting the results below, I will distinguish between effects that occur already in the large-modulation regime and those that rely on the formal regime in which the linearized hopping changes sign.

\subsection{Sign-problem-free quantum Monte Carlo and bipolaron properties}

The crucial methodological observation, made independently by my Columbia collaborators and I on one hand, and by the Amherst group of Prokof'ev and Svistunov and their postdoc Zhang in 2021 on the other, is that single polarons in the bond-Peierls model are also light --- this had been suspected from the site-Peierls work but had not been demonstrated in higher dimensions or in the adiabatic regime.\cite{ZhangProkofevSvistunovBondPolaron,CarboneMillisReichmanSousBondPolaron} Even more importantly, the model of Eq.~\eqref{eq:Hbond} has no sign problem in the singlet two-electron sector, and can therefore be solved numerically exactly using the path-integral / diagrammatic quantum Monte Carlo (QMC) method developed in Ref.~\cite{ZhangProkofevSvistunovQMC} (the absence of a sign problem is a special property of the bond version of the model: the site-Peierls model of Eq.~\eqref{eq:sitePeierls} does have a sign problem, because the coupling involves the difference of phonon displacements on neighboring sites, whereas the bond version involves a single oscillator whose sign can always be gauged away). This realization is what made the project I describe in this section --- estimates of $\Tc$ based on a a numerically exact computation of bipolarons in two dimensions --- possible.

Working with C. Zhang, D.~R. Reichman, M. Berciu, A.~J. Millis, N.~V. Prokof'ev and B.~V. Svistunov, we used this QMC approach to compute the bipolaron binding energy $\Delta_{\mathrm{BP}}$, the bipolaron effective mass $\mBP \equiv [(\partial^2 E_{\mathrm{BP}}(K)/\partial K^2)|_{K=0}]^{-1}$, and the bipolaron mean squared radius $R^2_{\mathrm{BP}} \equiv \langle\Psi_{\mathrm{BP}}|\hat{R}^2|\Psi_{\mathrm{BP}}\rangle$ as functions of $\lambda$, $t/\Om$ and $U/t$ on a $128\times 128$ square lattice.\cite{ZhangSousPRX2023} The technical details of the method are in Ref.~\cite{ZhangSousPRX2023} and in Ref.~\cite{ZhangProkofevSvistunovQMC}; the qualitative point relevant for this review is that, for the bond-Peierls model, the simulations are numerically exact, free of any sign problem, and converged on lattices large enough that finite-size effects are negligible. We find that, over a parametrically wide region of coupling and adiabaticity, the bipolaron mass remains within a factor of order unity of its non-interacting value $m_0 = 2m_{\mathrm{e}}$, the binding energy is large (of order $t$ at the optimal couplings), and the radial size remains of the order of one to a few lattice constants. The bipolaron is bound, small, and light --- exactly the combination needed for high-$\Tc$ bipolaronic superconductivity, and the combination that is forbidden in Holstein-type models.

\subsection{From bipolaron properties to $\Tc$}

In two dimensions a dilute gas of hard-core bosons of mass $\mBP$ and density $\nBP$ undergoes a Berezinskii--Kosterlitz--Thouless transition at\cite{FisherHohenberg,ProkofevRuebenackerSvistunov,PilatiGiorginiProkofev}
\begin{equation}
\Tc \;\approx\; 1.84\,\frac{\nBP}{\mBP}\bigg[1 + 0.29\,\ln\ln\frac{V_{\mathrm{BP\text{-}BP}}}{\nBP}\bigg]^{-1} \;\approx\; 1.84\,\frac{\nBP}{\mBP},
\label{eq:Tcformula}
\end{equation}
where the dependence on the bipolaron--bipolaron interaction strength $V_{\mathrm{BP\text{-}BP}}$ is only doubly logarithmic and can therefore be dropped to the accuracy of the present discussion. Equation~\eqref{eq:Tcformula} is valid as long as the bipolarons do not overlap, i.e.\ for $\nBP \lesssim 1/(\pi R^2_{\mathrm{BP}})$. The maximum $\Tc$ accessible from this mechanism is reached at the largest density at which the bipolarons remain non-overlapping, where they ``touch''; substituting $\nBP = \min\{1/(\pi R^2_{\mathrm{BP}}),\,1/\pi\}$ one obtains, for the case $R^2_{\mathrm{BP}} \geq 1$,
\begin{equation}
\Tc \;\approx\; \frac{0.5}{\mBP R^2_{\mathrm{BP}}},
\label{eq:Tcfinal}
\end{equation}
and for $R^2_{\mathrm{BP}}<1$ the saturation $\Tc \approx 0.5/\mBP$ takes over. Equation~\eqref{eq:Tcfinal} expresses the maximum $\Tc$ purely in terms of two ground-state bipolaron properties, $\mBP$ and $R^2_{\mathrm{BP}}$, that are accessible to QMC. The computation of $\Tc$ thus reduces, given the QMC results for $\mBP$ and $R^2_{\mathrm{BP}}$, to evaluating Eq.~\eqref{eq:Tcfinal}. I want to emphasize that  corrections from a more realistic treatment of the bipolaron--bipolaron interaction enter only at the doubly-logarithmic level, making this estimate reasonable. For a bipolaronic superconductor, the pairs must remain bound at the scale where superconducting coherence develops. Thus, the coherence scale estimated from Eq.~\eqref{eq:Tcfinal} must lie below the bipolaron binding energy $\Delta_{\mathrm{BP}}$. This condition is satisfied in all cases considered below, so $\Tc$ is controlled by the coherence scale.

\subsection{The result: bipolaronic high-$T_{\mathrm{c}}$ superconductivity}

\begin{figure}[t]
\raggedright
\centerline{\includegraphics[width=0.65\columnwidth]{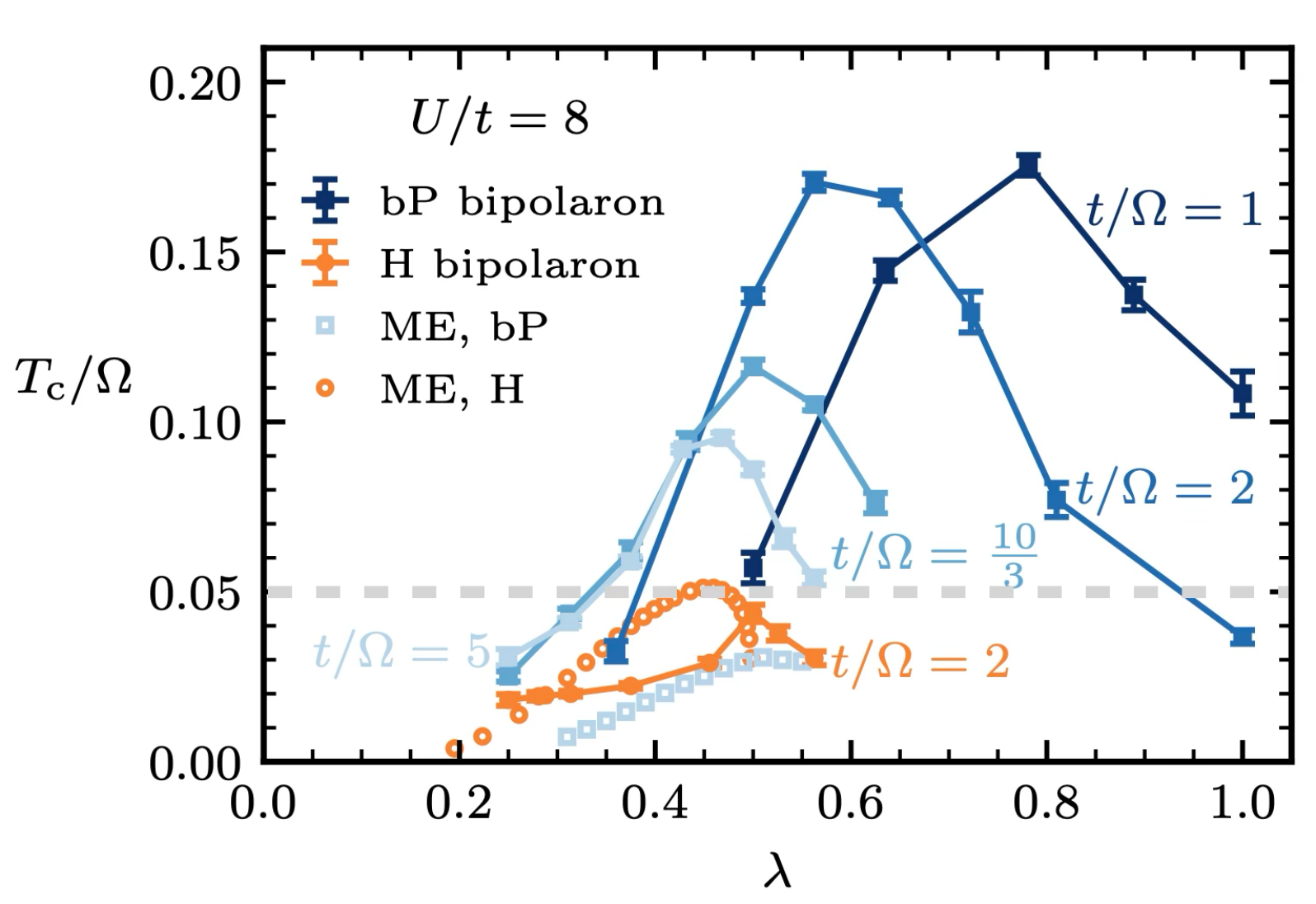}\hspace{1cm}}
\vspace{-2mm}
\caption{Bipolaronic high-$\Tc$ superconductivity in the bond-Peierls model in two dimensions, in units of the phonon frequency $\Om$, as a function of the dimensionless electron--phonon coupling $\lambda$ for several values of the adiabaticity ratio $t/\Om$ at on-site Hubbard repulsion $U = 8t$. Filled blue squares are QMC results for the bond-Peierls (bP) bipolaron computed from Eq.~\protect\eqref{eq:Tcfinal}; filled orange circles are the analogous result for the Holstein (H) bipolaron at $t/\Om = 2$; open symbols are Migdal--Eliashberg estimates for the same models. The dashed gray line at $\Tc/\Om \sim 0.05$ indicates the conventional weak-coupling/Migdal--Eliashberg bound. The bond-Peierls bipolaronic superconductor exceeds this ceiling over a broad region of parameter space. Reproduced from Ref.~\protect\cite{ZhangSousPRX2023}.}
\label{fig:PRXmain}
\end{figure}

Figure~\ref{fig:PRXmain} shows the result.\cite{ZhangSousPRX2023} The most striking feature is that, over a broad and physically relevant region of parameter space, $\Tc/\Om$ for the bond-Peierls bipolaronic superconductor lies well above the conventional ceiling indicated by the dashed gray line, and well above the Migdal--Eliashberg result for the same model. The maximum $\Tc/\Om$ in the calculation is of order $0.2$, reached for $t/\Om \sim 1$--$2$; for a phonon frequency $\Om \sim 200~\mathrm{K}$ this corresponds to a transition temperature in the range of several tens of kelvin, comparable to or exceeding the conventional ceiling at $30~\mathrm{K}$. The contrast with the Holstein bipolaronic superconductor (filled orange circles) is dramatic: in the Holstein model, $\Tc/\Om$ never exceeds $\sim 0.05$, and rapidly drops to vanishingly small values as the coupling is increased into the regime in which the bipolaron is well-defined. The contrast with Migdal--Eliashberg theory of either model (open symbols) is equally dramatic: ME theory generates a $\Tc/\Om$ that respects the conventional bound, by construction, and the bipolaronic estimate exceeds it cleanly.

The behavior of $\Tc/\Om$ as a function of $\lambda$ is non-monotonic and dome-shaped. This can be understood from the competition built into Eq.~\eqref{eq:Tcfinal}: as $\lambda$ is increased, the bipolaron becomes more strongly bound and shrinks (so $R^2_{\mathrm{BP}}$ decreases, helping $\Tc$), but it also becomes heavier (so $\mBP$ increases, hurting $\Tc$).

The two effects cross at an optimal $\lambda_{\mathrm{op}}$, beyond which the exponential mass enhancement that is the hallmark of the strong-coupling regime of typical electron--phonon models eventually wins. The peak of the dome shifts to smaller $\lambda$ as $t/\Om$ is increased, reflecting the fact that at fixed $\lambda$ a larger $t/\Om$ already places more phonons in the polaronic cloud and so brings on the mass enhancement earlier.

\subsection{Coulomb repulsion enhances $T_{\mathrm{c}}$}

An intriguing feature of the result is the response to a screened on-site Coulomb repulsion. Naively, a Hubbard $U$ should hurt an $s$-wave pairing state, since $s$-wave pairs have substantial on-site weight. In the Holstein model this expectation is borne out: turning on $U$ does eventually unbind the bipolaron, and although the mass decreases somewhat in the process, the simultaneous rapid growth of $R^2_{\mathrm{BP}}$ and the collapse of $\Delta_{\mathrm{BP}}$ overwhelm any benefit from the lighter mass, so $\Tc$ is monotonically suppressed.

\begin{figure}[t]
\centerline{\includegraphics[width=0.65\columnwidth]{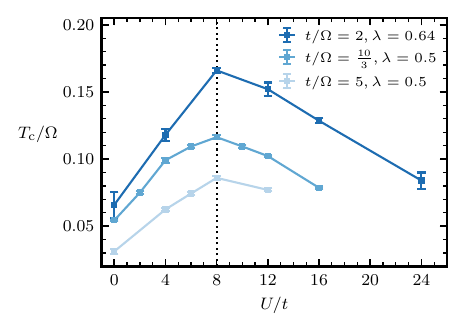}\hspace{1cm}}
\vspace{-4mm}
\caption{Coulomb-repulsion-mediated enhancement of bipolaronic high-$\Tc$ superconductivity in the bond-Peierls model. $\Tc/\Om$ for several values of $t/\Om$ at intermediate coupling $\lambda \sim 0.5$--$0.6$, as a function of the on-site Hubbard repulsion $U/t$. The transition temperature is enhanced as $U$ is turned on, and reaches a maximum near $U \approx 8t$ before decreasing. Reproduced from Ref.~\protect\cite{ZhangSousPRX2023}.}
\label{fig:Udome}
\end{figure}

The bond-Peierls model behaves in the opposite way. Figure~\ref{fig:Udome} shows that $\Tc/\Om$ is \emph{enhanced} by the Hubbard repulsion, exhibits a dome-like structure as a function of $U/t$, and is maximized near $U \approx 8t$ --- a value that is, strikingly, of the same order as the coupling found in transition-metal oxides. The microscopic origin of this behavior is visible in the QMC data for the bipolaron properties: at the relevant couplings the bipolaron has nontrivial weight on neighboring sites and not just on a single site, so it can avoid the on-site repulsion without sacrificing its binding. As $U$ is increased the mass $\mBP$ decreases (because the bipolaron is forced to spread out a little, and a more spread-out object is lighter in this model), while the squared radius $R^2_{\mathrm{BP}}$ depends only weakly on $U$ and the binding energy $\Delta_{\mathrm{BP}}$ decreases gradually rather than catastrophically. The combination drives $\Tc$ upward, until eventually the decreasing $\Delta_{\mathrm{BP}}$ becomes the bottleneck and $\Tc$ turns over. The result is an unusual phenomenon --- on-site Coulomb repulsion \emph{helping} an $s$-wave bipolaronic superconductor --- that is, to my knowledge, unique to the phonon-modulated-hopping family.

\section{Robustness against unscreened long-range Coulomb repulsion}\label{sec:bond3D}

The most natural objection to the analysis of the previous section is that the Coulomb repulsion in real materials is not a local Hubbard $U$ but a long-range $1/r$ tail, and that screening, although present, is by no means complete. The $\Tc$ scale in Eq.~\eqref{eq:Tcfinal} is small enough that even a modest unscreened component of the interaction could in principle wreck the bipolaronic mechanism, and the standard intuition --- larger couplings are required to bind two electrons against a long-range repulsion, larger couplings lead to heavier bipolarons, heavier bipolarons lead to lower $\Tc$ --- would suggest that this is indeed what happens. The question of whether this intuition is correct in the bond-Peierls model is the subject of Ref.~\cite{SousZhangPRB2023}, which I summarize here.

The model studied in Ref.~\cite{SousZhangPRB2023} is the three-dimensional version of Eq.~\eqref{eq:Hbond} on a cubic lattice, supplemented with both an on-site Hubbard $U$ and a long-range $1/r$ Coulomb tail:
\begin{equation}
\hat{V}_{\mathrm{Coulomb}} \;=\; U\sum_i \hat{n}_{i\uparrow}\hat{n}_{i\downarrow} + \frac{1}{2}\sum_{i\neq j}\frac{V}{|\mathbf{r}_i - \mathbf{r}_j|}\hat{n}_i\hat{n}_j.
\label{eq:Coulomb3D}
\end{equation}
The values $U = 8t$ and $V = U/10$ used in the simulations correspond, in physical units (with $V/|\mathbf{r}| \sim e^2/(\epsilon|\mathbf{r}|)$), to a dielectric constant of order $\epsilon \sim 10\,e^2/U \sim e^2/t$, comparable to that of transition-metal oxides. This is, by any reasonable standard, a substantial unscreened Coulomb repulsion, and it is the largest one accessible to the QMC approach.

The strategy is the same as in two dimensions: solve the two-electron problem exactly with QMC, extract $\mBP$ and $R^2_{\mathrm{BP}}$, and feed them into the appropriate three-dimensional analogue of Eq.~\eqref{eq:Tcfinal}. In three dimensions a dilute gas of hard-core bosons undergoes Bose--Einstein condensation at $\Tc \approx \#\,\nBP^{2/3}/\mBP$, with a numerical coefficient $\# \approx 3.2$--$3.3$ that has been determined to high precision from QMC simulations of the dilute Coulomb Bose gas in the same parameter regime.\cite{ZhangCapogrosso} The corresponding bound at the largest non-overlapping density gives a maximum $\Tc \approx 1.2/(\mBP R^2_{\mathrm{BP}})$, the three-dimensional analogue of Eq.~\eqref{eq:Tcfinal}. (The same QMC analysis confirms that, in the parameter regime relevant here, the dilute Coulomb Bose gas does not undergo Wigner crystallization and remains a superfluid.\cite{ZhangCapogrosso})

\begin{figure}[t]
\centerline{\includegraphics[width=0.675\columnwidth]{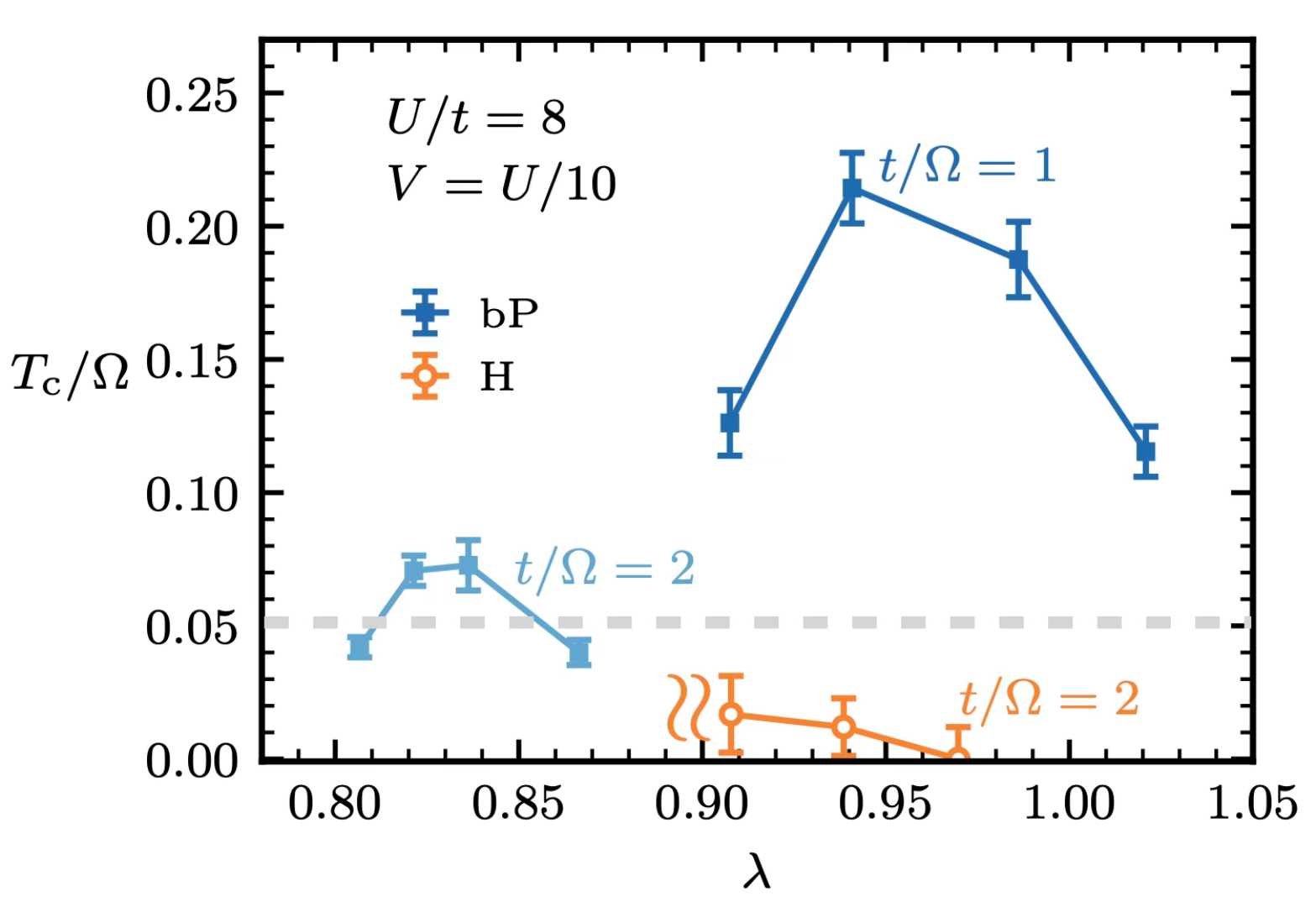}\hspace{1cm}}
\caption{Bipolaronic superconductivity in the bond-Peierls model in three dimensions, in the presence of an on-site Hubbard repulsion $U = 8t$ and an unscreened long-range Coulomb tail $V = U/10$, for several values of the adiabaticity ratio $t/\Om$ (filled blue squares: bond-Peierls; open orange circles: Holstein). The dashed gray line corresponds to the conventional weak-coupling ceiling at $\Tc/\Om \sim 0.05$. The peak $\Tc/\Om \sim 0.2$ at $t/\Om \approx 1$ corresponds to $\sim 20$~K for typical phonon frequencies. Reproduced from Ref.~\protect\cite{SousZhangPRB2023}.}
\label{fig:PRB108}
\end{figure}

The result is shown in Fig.~\ref{fig:PRB108}. Two qualitative observations are worth highlighting. First, the bipolaronic superconductor survives the unscreened long-range Coulomb repulsion essentially unscathed: $\Tc/\Om$ continues to exceed the conventional ceiling of $\sim 0.05$ over a parametrically wide region, and the maximum $\Tc/\Om \approx 0.2$ at $t/\Om \approx 1$ corresponds to a transition temperature of order $20$~K for typical phonon frequencies. Second, comparing Fig.~\ref{fig:PRB108} to Fig.~\ref{fig:PRXmain}, one sees that the long-range Coulomb tail does push the optimal $\lambda$ to larger values --- since binding now must overcome a longer-range repulsion --- and that the maximum $\Tc/\Om$ is somewhat reduced compared to the screened case. This is the price of working with an unscreened interaction, and screening (by gates, by a dielectric substrate, or by the material's own polarizability) would only help. The Holstein bipolaron, by contrast, is essentially absent from this region of parameter space; the orange line in Fig.~\ref{fig:PRB108} shows the very small $\Tc$ achievable in the Holstein model at $t/\Om = 2$ for the same $U$ and $V$, and indicates the absence of a well-defined bipolaron at smaller couplings, where a crossover to a BCS-like state may occur.

The take-home message is that the bipolaronic mechanism described here survives. It is not a fragile feature of an idealized lattice model with Hubbard interactions only; it persists despite the strong Coulomb repulsion present in real materials, including long-range tails of the same order of magnitude as those found in transition-metal oxides.

\section{Why are bipolarons light? A semi-classical instanton picture}\label{sec:semiclassical}

The QMC results of the previous two sections demonstrate that, in the bond-Peierls model, bipolarons are bound, small, and light over a wide region of parameter space, and that as a consequence the bipolaronic superconductor exceeds the conventional bound. They do not, however, by themselves explain \emph{why} bipolarons in phonon-modulated-hopping models are light, while bipolarons in density-coupled (Holstein) models are heavy. The pair-hopping argument of Sec.~\ref{sec:peierls} is suggestive but is restricted to the antiadiabatic limit; the QMC data extend over the full range of $t/\Om$ but are numerical. A controlled, asymptotic understanding valid in the opposite limit --- the adiabatic regime $t/\Om \gg 1$ relevant to most materials --- has been provided recently in Ref.~\cite{KimHanSousPRB2024} by K.~S. Kim, Z. Han, and the present author. I summarize the key idea here.

\begin{figure}[t]
\includegraphics[width=0.475\columnwidth,height=0.475\columnwidth]{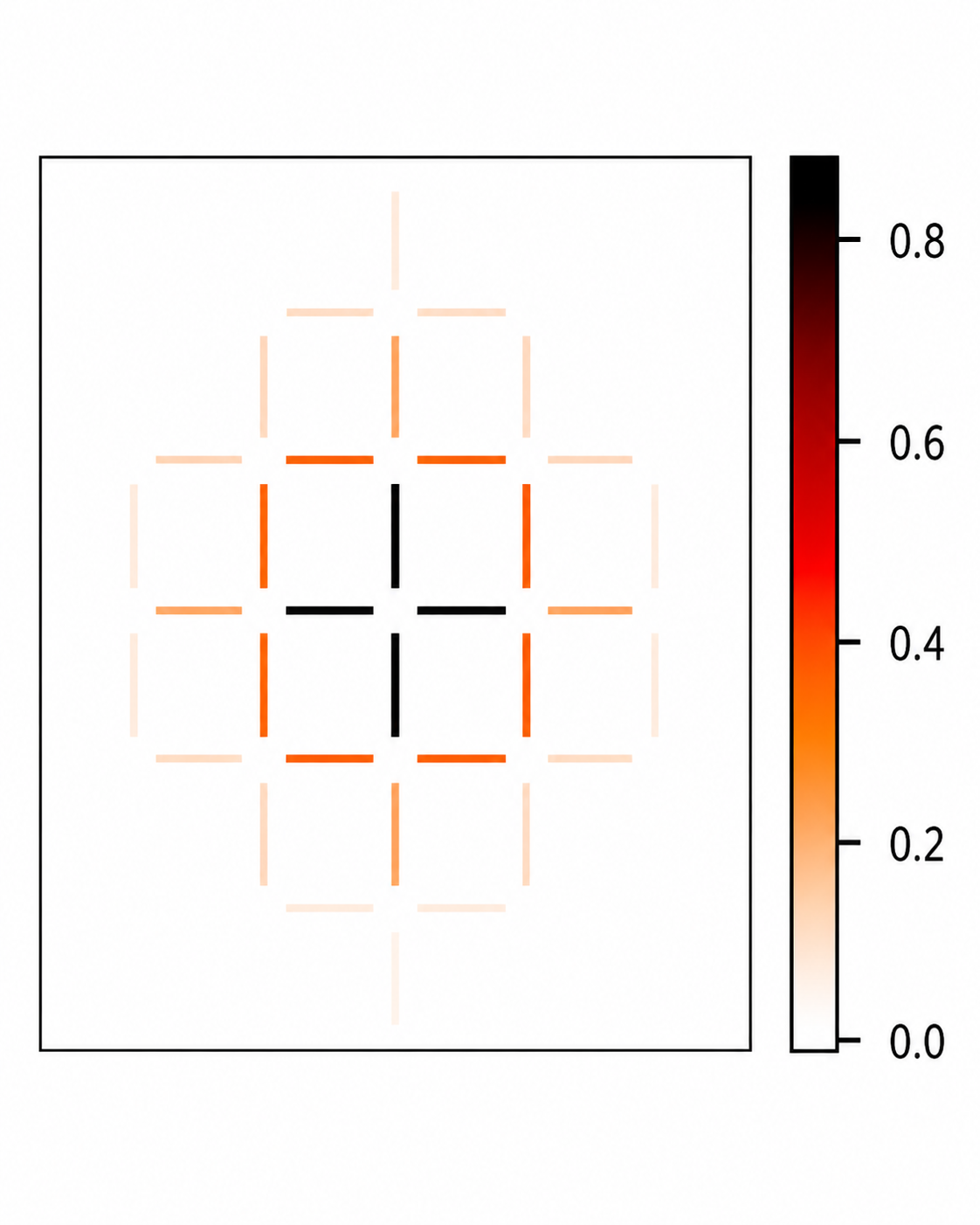}\hfill\includegraphics[width=0.475\columnwidth,height=0.475\columnwidth]{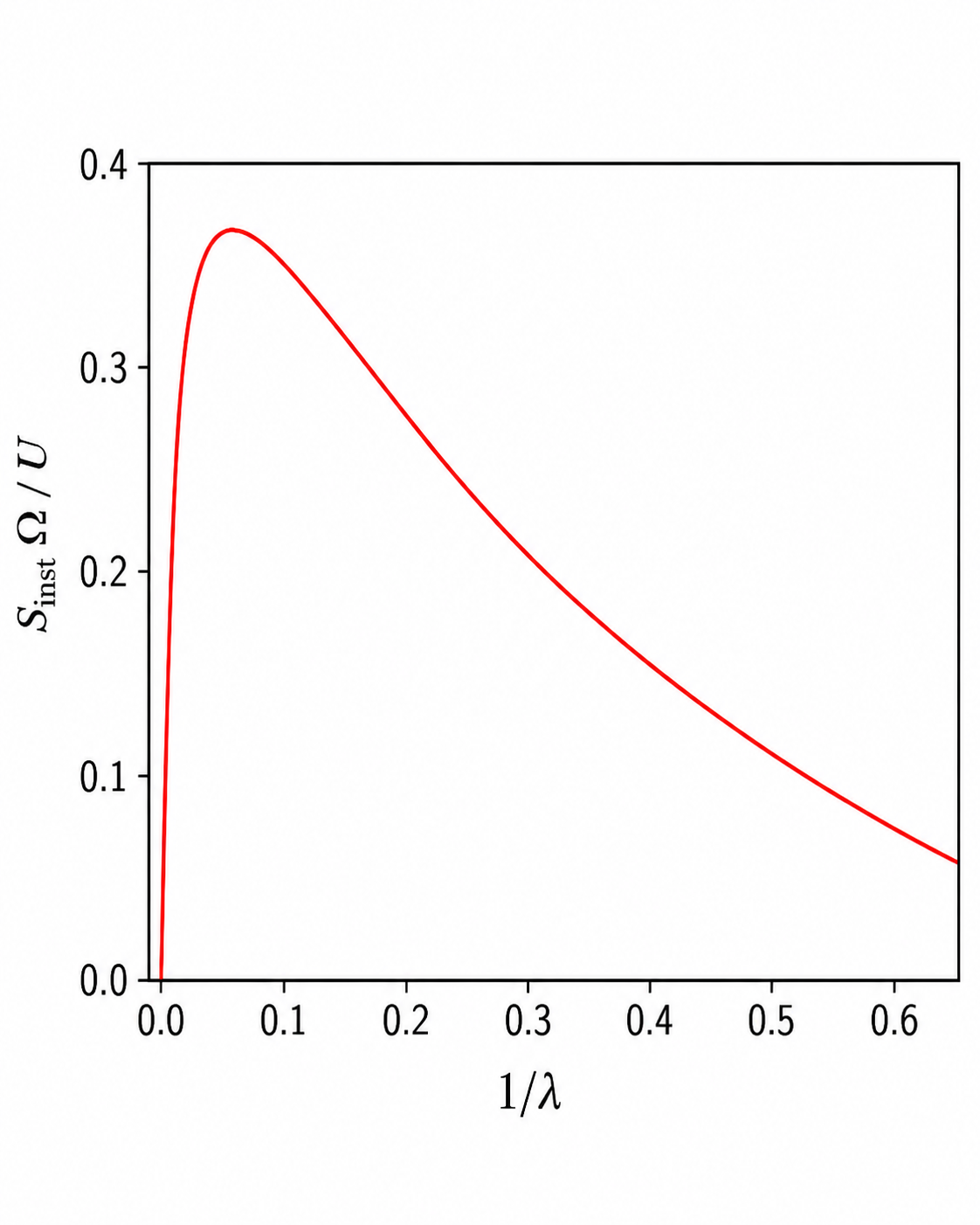}
\vspace{-6mm}
\caption{Left. Bipolaron configuration in the classical $\Om = 0$ limit of the bond-Peierls model: a cluster of neighboring bonds is distorted, with the electron pair localized on it. Right. Instanton action $S_{\mathrm{inst}}\Om/(t\lambda)$, which controls the bipolaron mass through quantum tunneling between near-degenerate classical configurations, as a function of $1/\lambda$. The barrier vanishes at strong coupling, leading to a light effective mass. Reproduced from Ref.~\protect\cite{KimHanSousPRB2024}. \emph{(Reuse from talk slide 39: ``Semi-classical theory of bipolaronic high-$T_{\mathrm{c}}$ superconductivity.'')}}
\label{fig:semiclassical}
\end{figure}

The starting point is the strong-coupling, classical-phonon limit $\Om \to 0$. In this limit the phonon coordinates are static variables, the electronic Hamiltonian becomes a quadratic problem in a frozen lattice background, and the bipolaron ground state is found by minimizing the total energy over both the electronic wavefunction and the phonon configuration. For the bond-Peierls model the result --- shown schematically as the heat-map in Fig.~\ref{fig:semiclassical}(a) --- is striking: the optimal phonon configuration consists of a few neighboring bonds whose lengths are reduced (and on which the electron pair lives), connected by a network of bonds with smaller distortions. Crucially, this optimal configuration has a continuous family of degenerate (or near-degenerate) translates, related by moving the cluster of distorted bonds across the lattice, and the energetic barrier between adjacent translates --- the height of the saddle the bipolaron must pass over to translate by one lattice spacing --- vanishes as $1/\lambda \to 0$.

Restoring the quantum dynamics of the phonons through a controlled instanton expansion around this classical solution, one finds that the bipolaron's effective mass $\mBP/m_0$ is set by the exponential of an instanton action $S_{\mathrm{inst}}$ that measures the cost of tunneling between adjacent classical configurations. For the bond-Peierls model the result is\cite{KimHanSousPRB2024}
\begin{equation}
\Tc \;\sim\; n\,\exp\!\left\{-\#\,\frac{t}{\Om}\sqrt{\lambda}\right\},
\label{eq:Tc_bP_asymptotic}
\end{equation}
in sharp contrast to the corresponding result for the Holstein (density-coupled) model,
\begin{equation}
\Tc \;\sim\; n\,\exp\!\left\{-\#\,\frac{t}{\Om}\,\lambda\right\}.
\label{eq:Tc_H_asymptotic}
\end{equation}
The two formulas differ, asymptotically at large $\lambda$, only in the exponent --- $\sqrt{\lambda}$ versus $\lambda$ --- but this difference is the difference between a $\Tc$ that is exponentially small but slowly varying in $\lambda$ and a $\Tc$ that is doubly exponentially suppressed. It is the asymptotic, quantitative statement of the qualitative observation that the bond-Peierls bipolaron is light and the Holstein bipolaron is heavy, and it is the explanation for the entire phenomenology of Sections~\ref{sec:bond2D} and~\ref{sec:bond3D}.

The physical content of Eqs.~\eqref{eq:Tc_bP_asymptotic} and~\eqref{eq:Tc_H_asymptotic} is that the bond-Peierls bipolaron, although small, lives in a phonon background whose energy landscape becomes flat at strong coupling --- the barrier the bipolaron must tunnel through in order to translate vanishes as $1/\lambda$, and the quantum dynamics of the phonons restores the translation symmetry that the classical configuration breaks. The Holstein bipolaron, by contrast, sits in a self-trapping well whose depth grows with $\lambda$, and translation requires the entire well to be reconstructed at the new site at exponential cost. The two scenarios are dynamically inequivalent at the deepest level, and the difference between them is what makes bipolaronic high-$\Tc$ superconductivity possible in the first case and impossible in the second.

\section{Salient features and outlook}\label{sec:outlook}

Let me collect the main messages of this review.

\emph{(i) Small, light bipolarons give relatively high $\Tc$.} The basic ingredient of the bipolaronic mechanism described here is the combination of a strong binding (small bipolaron size) with a relatively light effective mass. Both are simultaneously present in the bond-Peierls model and absent in Holstein-type density-coupled models, where strong binding implies an exponentially heavy bipolaron and light bipolarons exist only in the weakly bound limit.

\emph{(ii) The maximum $\Tc$ as a function of $\lambda$ is non-monotonic and dome-like.} Increasing $\lambda$ first enhances binding and shrinks the bipolaron, helping $\Tc$, before exponential mass enhancement eventually dominates and $\Tc$ decreases. The optimal $\lambda$ shifts to smaller values as $t/\Om$ is increased.

\emph{(iii) On-site Coulomb repulsion can enhance $\Tc$.} In the bond-Peierls model, an on-site Hubbard $U$ enhances $\Tc$ over a wide range, with a maximum near $U \approx 8t$, before suppressing it at very large $U$. This is the opposite of the behavior of Holstein bipolarons and reflects the fact that the bond-Peierls bipolaron has appreciable weight on neighboring sites and so can avoid the on-site repulsion without losing its binding.

\emph{(iv) Long-range Coulomb repulsion reduces but does not destroy bipolaronic high-$\Tc$ superconductivity.} An unscreened $1/r$ Coulomb tail of the magnitude found in transition-metal oxides shifts the optimal $\lambda$ to larger values and somewhat reduces the peak $\Tc/\Om$, but $\Tc$ remains substantially above the conventional ceiling. Screening helps.

\emph{(v) The maximum $\Tc$ in this mechanism exceeds the upper bound estimated from Migdal--Eliashberg theory.} This is the central result. The bipolaronic mechanism described here is, generically and over a wide region of parameter space, a more efficient route to phonon-mediated superconductivity than weak-coupling pairing out of a Fermi liquid.

\emph{(vi) The asymptotic mass-enhancement law is sub-exponential.} At strong coupling the bond-Peierls bipolaron has $\mBP \sim \exp(\#(t/\Om)\sqrt{\lambda})$, while the Holstein bipolaron has $\mBP \sim \exp(\#(t/\Om)\lambda)$. The square-root, rather than linear, dependence of the exponent on $\lambda$ is the asymptotic statement of the lightness of bipolarons in phonon-modulated-hopping models.

\medskip

I want to close with a few words about materials and design principles. The bond-Peierls coupling discussed here arises generically in any system where the orbitals of out-of-plane atoms mix with the bonding orbitals of in-plane atoms, so that transverse fluctuations of the out-of-plane atoms modulate the in-plane hopping. This is the situation in the iron-based pnictide superconductors,\cite{HauleNatMater,PnictogenHeight,GerberScience} where pnictogen oscillations modulate the Fe--Fe hopping through the destructive-interference mechanism shown in Fig.~\ref{fig:bondcartoon}; right panel. The relevant ratio $t/\Om \sim 2$--$3$ in FeSe places the system close to the regime in which the bipolaronic mechanism is most effective, and the dimensionless coupling $\lambda$ has been estimated to be of order $0.5$ in members of the family.\cite{ZXSHENFESE} The pairing symmetry of the iron-based superconductors has not been definitively established, but an ``extended $s$-wave'' state of the kind suggested by the bond-Peierls bipolaron is among the leading candidates. The bond-Peierls model is too simple to be quantitatively applied to the pnictides, which involve multi-orbital ``Hund's metal'' physics that is absent from Eq.~\eqref{eq:Hbond}, but the qualitative ingredients are arguably present, and a more realistic multi-orbital extension of the present analysis would be a natural next step.

A second class of materials where related physics may be operative are the cuprates, in which buckled Cu--O bonds and the associated apical-oxygen modes have long been argued to provide a phonon-mediated contribution to pairing.\cite{ZXNagaosaPeierlsCuprates,90bond,corner-sharing} Here the situation is more complicated, since cuprate superconductivity is widely (and in my view correctly) attributed primarily to electronic correlations, but the bond-Peierls mode may still contribute as a secondary channel that enhances $d$-wave pairing.

From a design standpoint, the analysis above suggests three concrete strategies for engineering the bipolaronic mechanism into novel materials. First, one wants to be in the ``quantal'' regime $t/\Om \sim 1$--$2$ in which phonons and electrons are competitive energetically; this is unusual for ordinary solids, where typically $t \gg \Om$, but it can be approached either by softening the lattice (for example by structural or moir\'e engineering, as in twisted bilayer graphene\cite{MAG,MoireFunctional}) or by stiffening the phonons (for example by incorporating light atoms on the bonds, as in superatomic crystals\cite{SuperatomicMaterials}). Second, the relevant electron--phonon coupling is the bond-modulating one, and this calls for materials with bridging atoms whose displacement modulates the dominant hopping pathway --- the pnictide architecture is a natural template, but other geometries should be explored. Third, the analysis of Sec.~\ref{sec:bond2D} shows that a moderate on-site Coulomb repulsion is not only acceptable but actively beneficial, while long-range Coulomb tails are tolerated rather than welcomed; gating, dielectric encapsulation, or substrate-induced screening are useful tools.

There are several open theoretical directions that the work I have described leaves untouched. The model studies discussed above stop at densities small enough that bipolarons do not overlap, and the question of what happens at larger densities --- whether the bipolaronic liquid evolves smoothly into a denser, more strongly correlated superconducting state, or whether competing instabilities take over --- requires methods that go beyond exact two-electron QMC. Whether the mechanism operates in models with both bond and site couplings active simultaneously (as is generically the case in real materials) is a natural and pressing question. And the extension of the present analysis to the multi-orbital, Hund's-coupled setting of the pnictides would close the gap between the schematic model studied here and a serious materials description. Beyond phonon-mediated superconductivity, the same notion --- that bound pairs of electrons can be small and light if their binding is generated by a kinetic-energy-enhancing rather than potential-energy-lowering interaction --- may be useful in other contexts, including light-induced superconductivity driven by photon-mediated couplings and the magnetically induced superconductivity reported in aromatic hydrocarbons. I will leave these directions for future work.

\section*{Acknowledgments}

I am deeply grateful to the organizers of the Athens Workshop in Theoretical Physics: 10th Anniversary, held at the National and Kapodistrian University of Athens on December 17--19, 2025, for the invitation to give a talk based on the work reviewed here, and for the opportunity to write it up as the present persective.
I am especially indebted to the mentorship and friendship of M.~Berciu, S.~A. Kivelson, R.~V. Krems, A.~J. Millis, and D.~R. Reichman. The work described here would not have been possible without the contributions of my collaborators: A.~J. Millis and D.~R. Reichman (Columbia), M. Berciu and R.~V. Krems (UBC), C. Zhang (East China Normal University and University of Science and Technology of China), N.~V. Prokof'ev and B.~V. Svistunov (UMass Amherst), K.~S. Kim (UIUC), Z. Han (Harvard), M. Chakraborty (Chennai Institute of Technology), and M. Carbone (Palantir). 

Financial support is acknowledged as follows. J.S.\ acknowledges support from Yale University and the Energy Sciences Institute at Yale. Earlier portions of this work were supported by the Gordon and Betty Moore Foundation's EPiQS Initiative through Grant GBMF8686 at Stanford University, the National Science Foundation Materials Research Science and Engineering Centers (MRSEC) program through Columbia University in the Center for Precision Assembly of Superstratic and Superatomic Solids under Grant No.\ DMR-1420634, by the Natural Sciences and Engineering Research Council of Canada (NSERC), and the Stewart Blusson Quantum Matter Institute.


\end{document}